\documentclass[floatfix,reprint,prl,superscriptaddress,twocolumn]{revtex4}
\usepackage{graphicx}
\usepackage{amsmath}
\usepackage{amsfonts}
\usepackage{amssymb}
\usepackage{amsbsy}
\usepackage{hyperref}

\begin{document}
\title{Electric field control of the magnetic chiralities in ferroaxial multiferroic RbFe(MoO$_4$)$_2$}
\date{\today}

\author{Alexander J. Hearmon}
\email{a.hearmon@physics.ox.ac.uk}
\affiliation{Clarendon Laboratory, Department of Physics, University of Oxford, Parks Road, Oxford OX1 3PU, UK}
\affiliation{Diamond Light Source Ltd., Harwell Science and Innovation Campus, Didcot, OX11 0DE, UK}
\author{Federica Fabrizi}
\affiliation{Clarendon Laboratory, Department of Physics, University of Oxford, Parks Road, Oxford OX1 3PU, UK}
\author{Laurent C. Chapon}
\affiliation{ISIS Facility, STFC-Rutherford Appleton Laboratory, Didcot, OX11 OQX, UK}
\affiliation{Institute Laue Langevin, 6 rue Jules Horowitz, 38042 Grenoble Cedex 9, France}
\author{R. D. Johnson}
\affiliation{Clarendon Laboratory, Department of Physics, University of Oxford, Parks Road, Oxford OX1 3PU, UK}
\affiliation{ISIS Facility, STFC-Rutherford Appleton Laboratory, Didcot, OX11 OQX, UK}
\author{Dharmalingam Prabhakaran}
\affiliation{Clarendon Laboratory, Department of Physics, University of Oxford, Parks Road, Oxford OX1 3PU, UK}
\author{Sergey V. Streltsov}
\affiliation{Institute of Metal Physics, Ural Division, Russian Academy of Sciences, ul. S. Kovalevsko\u{\i} 18, Yekaterinburg, 620041 Russia}
\affiliation{Ural Federal University, Mira St. 19, 620002 Ekaterinburg, Russia}
\author{P. J. Brown}
\affiliation{Institute Laue Langevin, 6 rue Jules Horowitz, 38042 Grenoble Cedex 9, France}
\author{Paolo G. Radaelli}
\affiliation{Clarendon Laboratory, Department of Physics, University of Oxford, Parks Road, Oxford OX1 3PU, UK}

\begin{abstract}
The coupling of magnetic chiralities to the ferroelectric polarisation in multiferroic RbFe(MoO$_4$)$_2$ is investigated by neutron spherical polarimetry. Because of the axiality of the crystal structure below $T_\textrm{c}$ = 190 K, helicity and triangular chirality are symmetric-exchange coupled, explaining the onset of the ferroelectricity in this proper-screw magnetic structure --- a mechanism that can be generalised to other systems with ``ferroaxial'' distortions in the crystal structure. With an applied electric field we demonstrate control of the chiralities in both structural domains simultaneously.  
\end{abstract}
\maketitle

Multiferroic materials, in which ferroelectricity and magnetic order coexist, are attracting conspicuous interest as candidates for novel applications in digital storage devices \cite{cheong2007multiferroics_8}. The requirement of having a strong magneto-electric effect has focussed the research on compounds in which ferroelectricity appears as a consequence of magnetic ordering (so called `type-II multiferroics'), leading to the realisation that cycloidal multiferroics, such as TbMnO$_3$ \cite{kimura2003magnetic_231} and Ni$_3$V$_2$O$_8$ \cite{Kenzelmann2006}, exhibit an exceptionally strong cross-coupling between the different types of order. In these compounds the atomic spins $\mathbf{S}_i$  rotate within a plane that contains the propagation direction of the incommensurate modulation $\hat{\mathbf{r}}_{i,i+1}$ which connects neighbouring atoms $\mathbf{S}_i$ and $\mathbf{S}_{i+1}$.
 
A phenomenological theory based on symmetry analysis of the magnetic and ferroelectric order parameters established that the magnetoelectric coupling has a trilinear form $\mathbf{P}\cdot[\mathbf{M}(\nabla \cdot \mathbf{M}) - (\mathbf{M}\cdot\nabla)\mathbf{M}]$ in the free energy \cite{Lifshitz}, leading to a polarisation given by $\mathbf{P} \propto \hat{\mathbf{r}}_{i,i+1} \times (\mathbf{S}_i \times \mathbf{S}_{i+1})$ \cite{mostovoy2006ferro_113}. A number of microscopic theories have been proposed that respect these symmetry constraints for the canonical multiferroic materials \cite{Sergienko2006role_204,katsura2005spin}.

An intense experimental effort and a refined understanding of the symmetry requirements for spin-driven ferroelectricity have expanded the range of candidate multiferroic materials. An interesting line of research has developed specifically on magnetochiral or proper-screw systems, in which the atomic spins rotate perpendicularly to the propagation direction of the screw \cite{arima2007ferroelectricity_115}; the model in ref. \onlinecite{mostovoy2006ferro_113} predicts no polarisation in this case. One mechanism that can lead to electrical polarisation in a proper-screw magnetic structure is the coupling to \emph{structural axiality} (ferroaxial coupling) \cite{Johnson2011}. A crystal structure can be considered `axial' if there exists a structural distortion that is unchanged by inversion and makes the two senses of rotation (clockwise or counterclockwise) about  an axial vector $\mathbf{A}$ distinguishable. As pointed out in ref. \onlinecite{Johnson2011}, the magnetic helicity $\sigma$ can be coupled phenomenologically to the electric polarisation $\mathbf{P}$ and structural axiality $\mathbf{A}$ to create a trilinear form, $\sigma \mathbf{A} \cdot \mathbf{P}$, which is invariant under spacial inversion and time reversal.  Very recently, ``giant'' magnetically-induced ferroelectricity has been reported in the ferroaxial system CaMn$_7$O$_{12}$ \cite{JohnsonCaMn7O12}.

RbFe(MoO$_4$)$_2$ (RFMO, ferroaxial below 190 K) is an extremely interesting system to test the interplay between ferroaxiality, magnetism and ferroelectricity. Below $T_\textrm{N} \approx 4$ K, it orders magnetically in a complex structure that possesses both helicity and triangular chirality (see below), becoming ferroelectric at the same temperature \cite{kenzelmann2007direct_160}. Yet, its layered crystal structure and exchange pathways are sufficiently simple for it to be considered the ``hydrogen atom'' of ferroaxial multiferroics.  In previous work, it has been suggested that the 120$^\circ$ magnetic structure in each layer is in itself sufficient to break inversion \cite{kenzelmann2007direct_160,plumer1991chirality_267}. Here, we demonstrate that the axiality and magnetic helicity also play a crucial role in the onset of ferroelectricity. We show that the helicity and triangular chirality are in fact coupled together in the free energy by the axial distortion, and we present data that show switching of both parameters simultaneously with an applied electric field.

\begin{figure}[t]
\begin{center}
\includegraphics[scale=0.5]{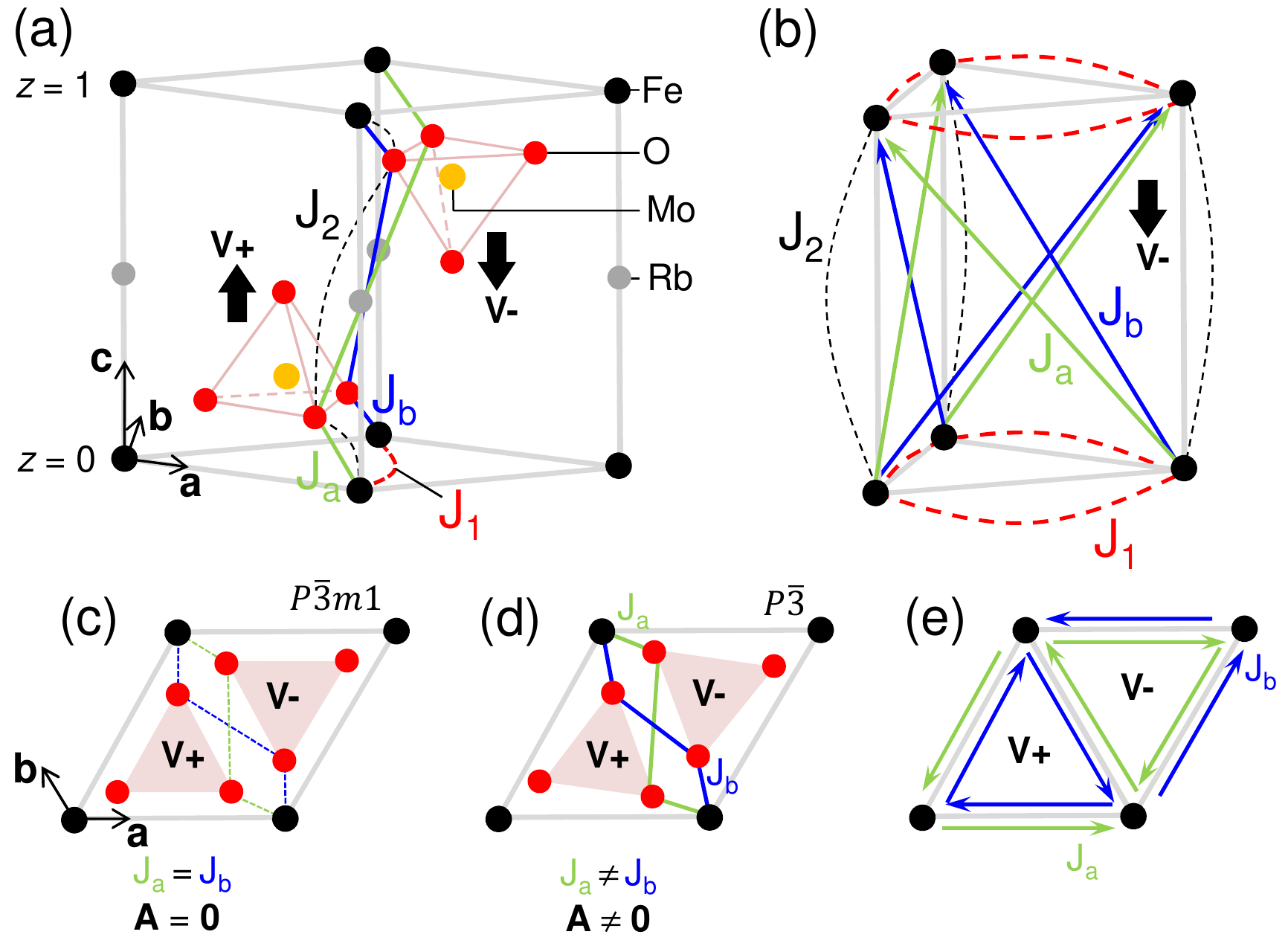}
\caption{(Color online) The structure and exchange paths of RFMO. (a) 3D view showing in-plane ($J_1$) and vertical ($J_2$) paths, as well as the two diagonal paths ($J_\textrm{a}$ and $J_\textrm{b}$) which depend on the ferroaxial distortion. The directions of the MoO$_4$ tetrahedra ($\mathbf{v}^+$ or $\mathbf{v}^-$) are indicated. (b) Shows the effect of applying the 3-fold symmetry on the interactions. (c)-(e) Illustrate the difference between the $J_\textrm{a}$ and $J_\textrm{b}$ paths arising from the distortion parametrized by the axial vector $\mathbf{A}$. (Arrows point from $z = 0$ to $z = 1$.) \label{Fig_structure}}
\end{center}
\end{figure}

RFMO undergoes a structural transition at $T_\textrm{c} = 190$ K in which the MoO$_4$ tetrahedra rotate (Fig. \ref{Fig_structure}), lowering the symmetry from $P\bar{3}m1$ to $P\bar{3}$ (ferroaxial point group $\bar{3}$) \cite{inami2007neutron_116,waskowska2010temp_159}. Below $T_\textrm{N}$ the Fe spins (one per unit cell) order magnetically in the $ab$ plane in a 120$^\circ$ structure, and rotate in a helix between one plane and the next with an incommensurate propagation along $c^*$ ($q_z \approx 0.44$, here always chosen to be \emph{positive}).  Fig. \ref{Fig_structure}(a) shows a unit cell of RFMO with the location of the MoO$_4$ tetrahedra indicated (oxygen atoms mediate the relevant exchange paths --- see below).  One can see that there is a MoO$_4$ tetrahedron above (or below) each magnetic triangle.  We define triangles as ``positive'' (``negative'') if they are associated with a tetrahedron pointing along the positive (negative) $c^*$ direction [this is indicated by the two unit vectors  $\mathbf{v}^+$ and $\mathbf{v}^-$ in fig. \ref{Fig_structure}(a)].  In plan view, one can rotate the structure in such a way that all the ``positive'' triangles point ``up'' the page [this orientation is depicted in Figs.\ref{Fig_structure}(c)--\ref{Fig_structure}(e)].  We define a \emph{staggered triangular chirality} (a \emph{parity odd} quantity) as $\sigma_\textrm{t}=\left(\mathbf{S}_1  \times \mathbf{S}_2+  \mathbf{S}_2  \times \mathbf{S}_3 + \mathbf{S}_3  \times \mathbf{S}_1\right) \cdot \mathbf{v}/S^2$.  In other words, $\sigma_\textrm{t} = 1$ if the spins rotate \emph{counterclockwise} as one circumscribes \emph{counterclockwise} a ``positive'' triangle of Fe spins (with a $\mathbf{v}^+$ tetrahedron at its centre); $\sigma_\textrm{t} = -1$ otherwise.  The magnetic \emph{helicity} is defined as usual as $\sigma_\textrm{h}=(\mathbf{S}_{z=0} \times \mathbf{S}_{z=1}) \cdot \hat{\mathbf{r}}_{01}/S^2 = \pm 1$ for a right- and left-handed magnetic screw, respectively. There are two possible magnetic propagation vectors:  $\mathbf{q}_1 = (1/3, 1/3, q_z)$ for $\sigma_\textrm{t} \sigma_\textrm{h}=1$ and $\mathbf{q}_2 = (-1/3,-1/3,q_z)$ for $\sigma_\textrm{t} \sigma_\textrm{h}=-1$, leading to distinct sets of peaks in the magnetic scattering for the two different combinations of chiralities.

Fig. \ref{Fig_structure}(a) shows a unit cell of RFMO with the location of four exchange paths (mediated by the MoO$_4$ oxygens) indicated. The super-superexchange terms across the prismatic faces, $J_\textrm{a}$ and $J_\textrm{b}$, are equal by symmetry in $P\bar{3}m1$ but become distinct in $P\bar{3}$, so that the difference ($J_\textrm{a} - J_\textrm{b})$ is proportional to the amplitude and sign of the ferroaxial distortion [Figs.\ref{Fig_structure}(c) and \ref{Fig_structure}(d)]. Figs. \ref{Fig_structure}(b) and \ref{Fig_structure}(e) show the effect of the ferroaxial rotation on the exchange paths. Assuming the 120$^\circ$ structure as given, the symmetric-exchange energy per Fe ion is   
\begin{equation}
E = E_0 + (\sqrt{3}/2)\sigma_{\textrm{h}}\sigma_{\textrm{t}}(J_\textrm{a}-J_\textrm{b})S^2\sin(2\pi q_z)
\label{energy}\end{equation}  
where $E_0 = -3J_1 S^2/2 + (J_2 - J_\textrm{a}/2 - J_\textrm{b}/2) S^2 \cos(2\pi q_z)$ doesn't depend on the magnetic chiralities. Minimising with respect to $q_z$ results in
\begin{equation}
\tan(2\pi q_z) = \sqrt{3} \sigma_\textrm{t} \sigma_\textrm{h}(J_\textrm{a} - J_\textrm{b})/(2J_2 - [J_\textrm{a} + J_\textrm{b}]). \label{prop_vec}
\end{equation}
We arrive here at our first important result:  in the presence of a ferroaxial crystal structure, adjacent 120$^\circ$ triangular magnetic planes will rotate with respect to each other, forming a helix, \emph{without} the need for antisymmetric or inter planar next-nearest neighbour interactions.  In a real crystal,  both ferroaxial domains will be present with roughly equal populations (one with $J_\textrm{a} > J_\textrm{b}$, the other with $J_\textrm{b} > J_\textrm{a}$).  Therefore, the lowest energy magnetic configurations will depend on the `axiality' of the domain in question. Notating the chiralities of a particular magnetic structure as $(\sigma_\textrm{t},\sigma_\textrm{h}) = (\pm, \pm)$, eqn. (\ref{energy}) implies that $\{(+,+),(-,-)\}$ states will be lower in energy than $\{(+,-), (-,+)\}$ if $J_\textrm{b} > J_\textrm{a}$, or vice versa for $J_\textrm{b} < J_\textrm{a}$.   One important consequence is that each of the two \emph{structural} axial domains will order with a distinct \emph{magnetic} propagation vector, say $\mathbf{q}_1$ for positive axiality and $\mathbf{q}_2$ for negative axiality (the exact combination depends on the sign of the magneto-elastic interaction).  Therefore, each set of distinct magnetic peaks ($\mathbf{q}_1$ or $\mathbf{q}_2$ peaks) probes a \emph{single} axial domain. 

\begin{figure}[t]
\begin{center}
\includegraphics[scale=0.5]{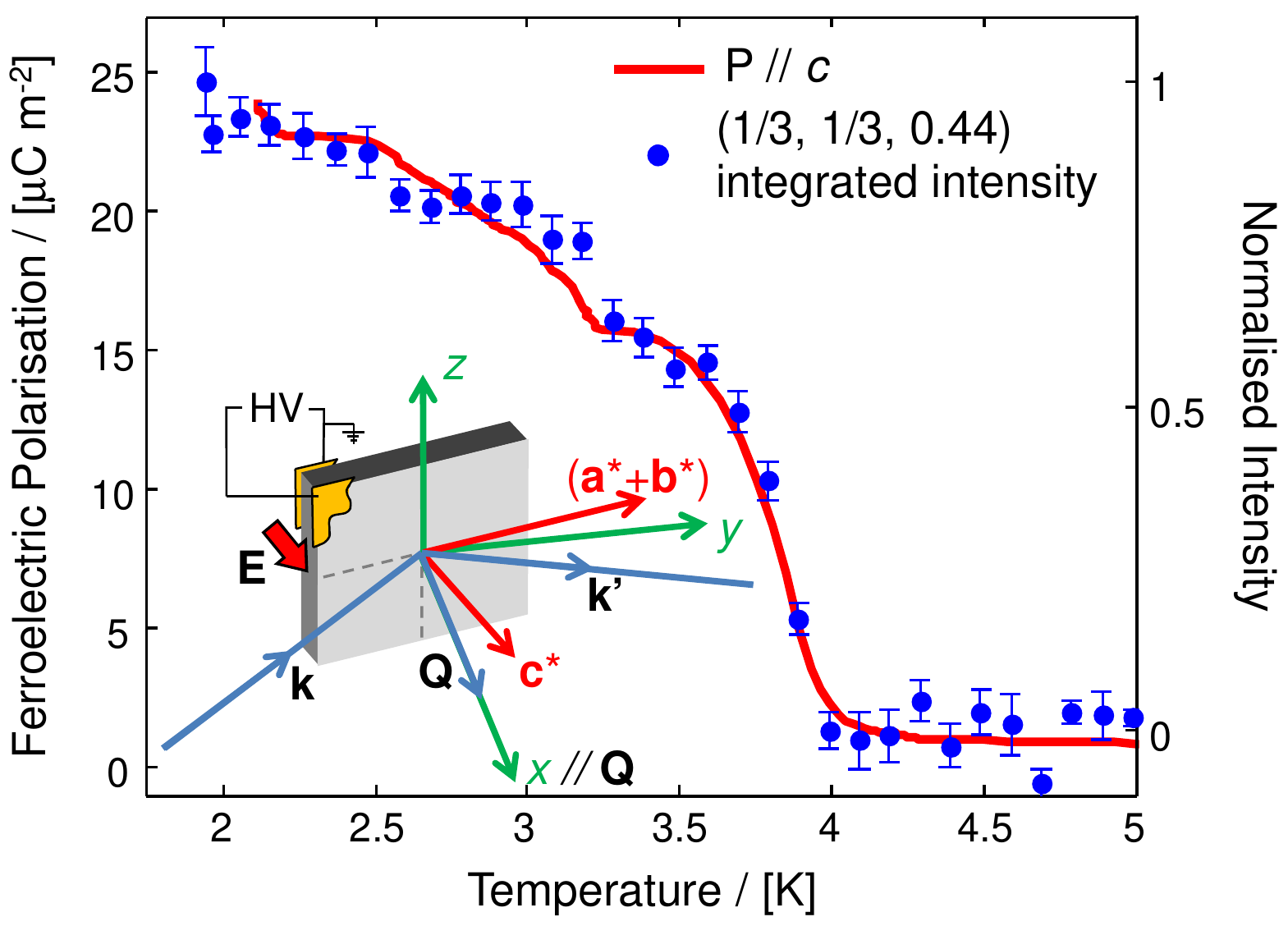}
\caption{(Color online) $c$-axis polarisation (obtained by integrating the pyroelectic current) and intensity of the $(1/3, 1/3, q_z)$ magnetic reflection (measured without polarisation analysis) as a function of temperature. Inset: the experimental geometry showing the crystal orientation, direction of applied $\mathbf{E}$-field, and scattering vector  $\mathbf{Q} = \mathbf{k}^\prime - \mathbf{k}$. \label{Fig_Temp_dep}}
\end{center}
\end{figure}

Spherical neutron polarimetry is an ideal technique to study these magnetic structures and their relationship with the electrical polarisation, since it has the ability to distinguish domains with different chiralities. In the present case the magnetic peaks do not overlap with the nuclear peaks in reciprocal space, so it is possible to calculate the polarisation of the scattered beam purely from the magnetic structure factor $\mathbf{M}$ \cite{blume1963polarization_191}. For this we work in the Blume reference frame, in which the $X$-axis is along the scattering vector $\mathbf{Q}$, the $Z$-axis is vertical, and the $Y$-axis completes the right-handed set (see inset to Fig. \ref{Fig_Temp_dep}). Magnetic neutron diffraction is only sensitive to the component of $\mathbf{M}$ perpendicular to $\mathbf{Q}$, which we write $\mathbf{M}_\perp = (0,M_{\perp y},M_{\perp z})$. For a structure with one magnetic ion per unit cell, the $l$th moment (in the unit cell of the lattice vector $\mathbf{R}_l$) is given by $\boldsymbol{\mu}_l = \mu_{l0}(\hat{\mathbf{u}} \mp i\hat{\mathbf{v}})\exp(i\mathbf{q}_1\cdot\mathbf{R}_l) + \textrm{c.c.}$ for $(+,+)$ or $(-,-)$ structures; and is given by $\boldsymbol{\mu}_l = \mu_{l0}(\hat{\mathbf{u}} \pm i\hat{\mathbf{v}})\exp(i\mathbf{q}_2\cdot\mathbf{R}_l) + \textrm{c.c.}$ for $(+,-)$ or $(-,+)$, where $\hat{\mathbf{u}}$ and $\hat{\mathbf{v}}$ are orthogonal unit vectors in the plane of the spins. Thus the magnetic structure factor $\mathbf{M}(\mathbf{Q}) = p f^{\textrm{mag}}(\mathbf{Q})(\hat{\mathbf{u}} \pm i\hat{\mathbf{v}})$ where $p$ is a constant, $f^{\textrm{mag}}$ is the magnetic form factor, and the $\pm$ is determined by the magnetic chiralities (see table \ref{peaks}). Given a fully polarised incident beam directed along $i$, the polarisation measured along $j$ is written $P_{ij}$ ($i,j = X,Y,Z$). For our geometry, the equations reduce to $P_{xx} = -1; P_{xy} = P_{xz} = P_{zy} = P_{yz} = 0;$ and
\begin{eqnarray} 
P_{yy} &=& 2M_{\perp y}^2/M_\perp^2 - 1; \label{pol1} \\
P_{zz} &=& 2M_{\perp z}^2/M_\perp^2 - 1; \\ 
P_{yx} &=& P_{zx} = 2\Im(M_{\perp y}M_{\perp z}^*)/M_\perp^2, \label{pol2}
\end{eqnarray} 
where $M_{\perp}^2 = \mathbf{M}_\perp \cdot \mathbf{M}_\perp^*$ and $M_{\perp i}^2 = M_{\perp i}M_{\perp i}^*$. Thus, the sign of the $P_{yx}$ and $P_{zx}$ elements, together with the position of the satellites ($\pm \mathbf{q}_{1,2}$) \emph{uniquely} determines the domain population for each of the $(\sigma_\textrm{t},\sigma_\textrm{h})$.

\begin{figure}[b]
\begin{center}
\includegraphics[scale=0.5]{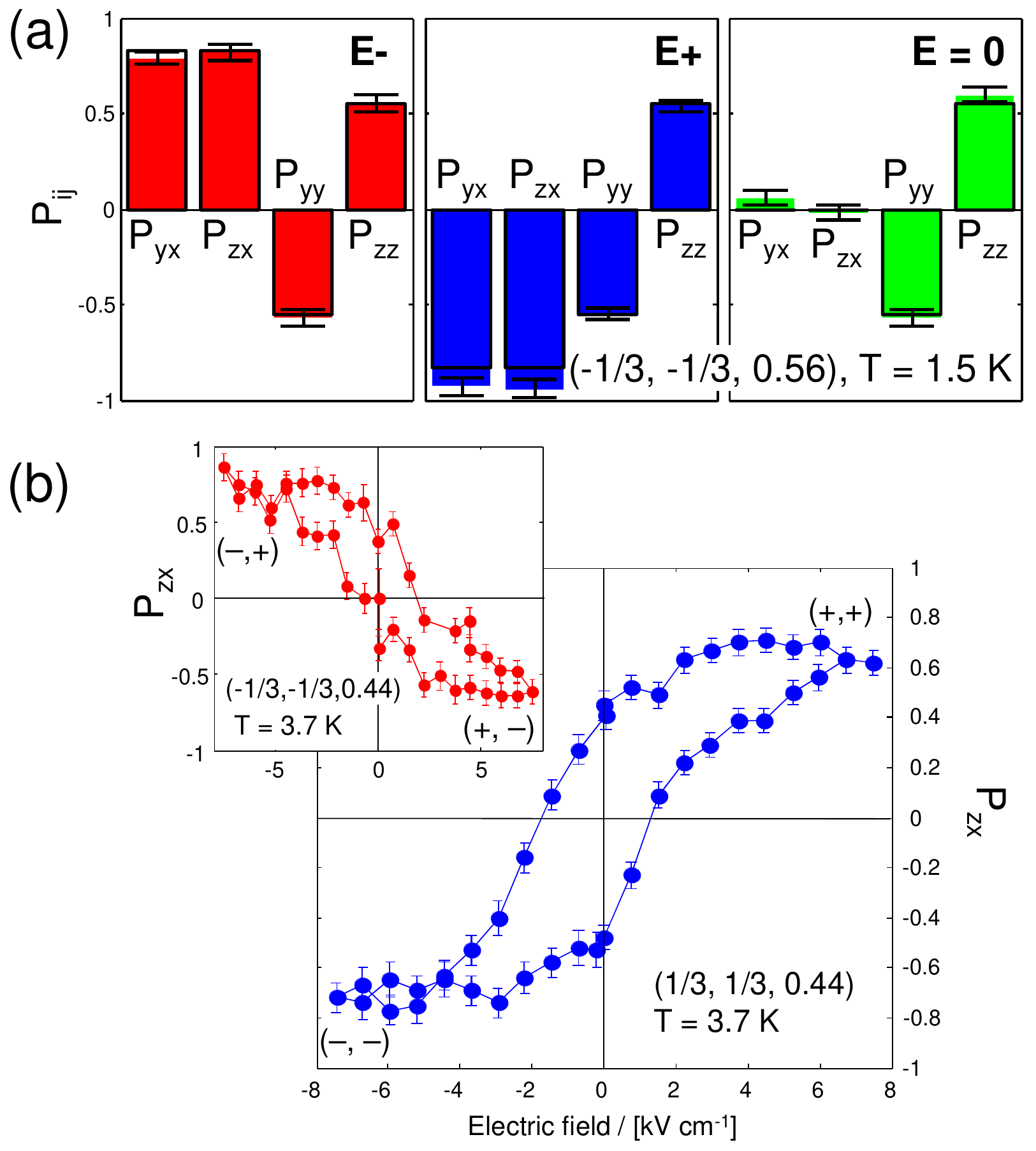}
\caption{(Color online) (a) Polarimetry components following negative, positive, and zero field cooling (with field strength 7.5 kV/cm). Solid bars indicate observed values and rectangles show the calculations. (b) Hysteresis loops in $P_{zx}$ as a function of applied electric field for two magnetic reflections. The chiralities, $(\sigma_\textrm{t},\sigma_\textrm{h}) = (\pm,\pm)$ are indicated.  \label{Fig_Bars_Hyst}}
\end{center}
\end{figure}

\begin{figure}[b]
\begin{center}
\includegraphics[scale=0.35]{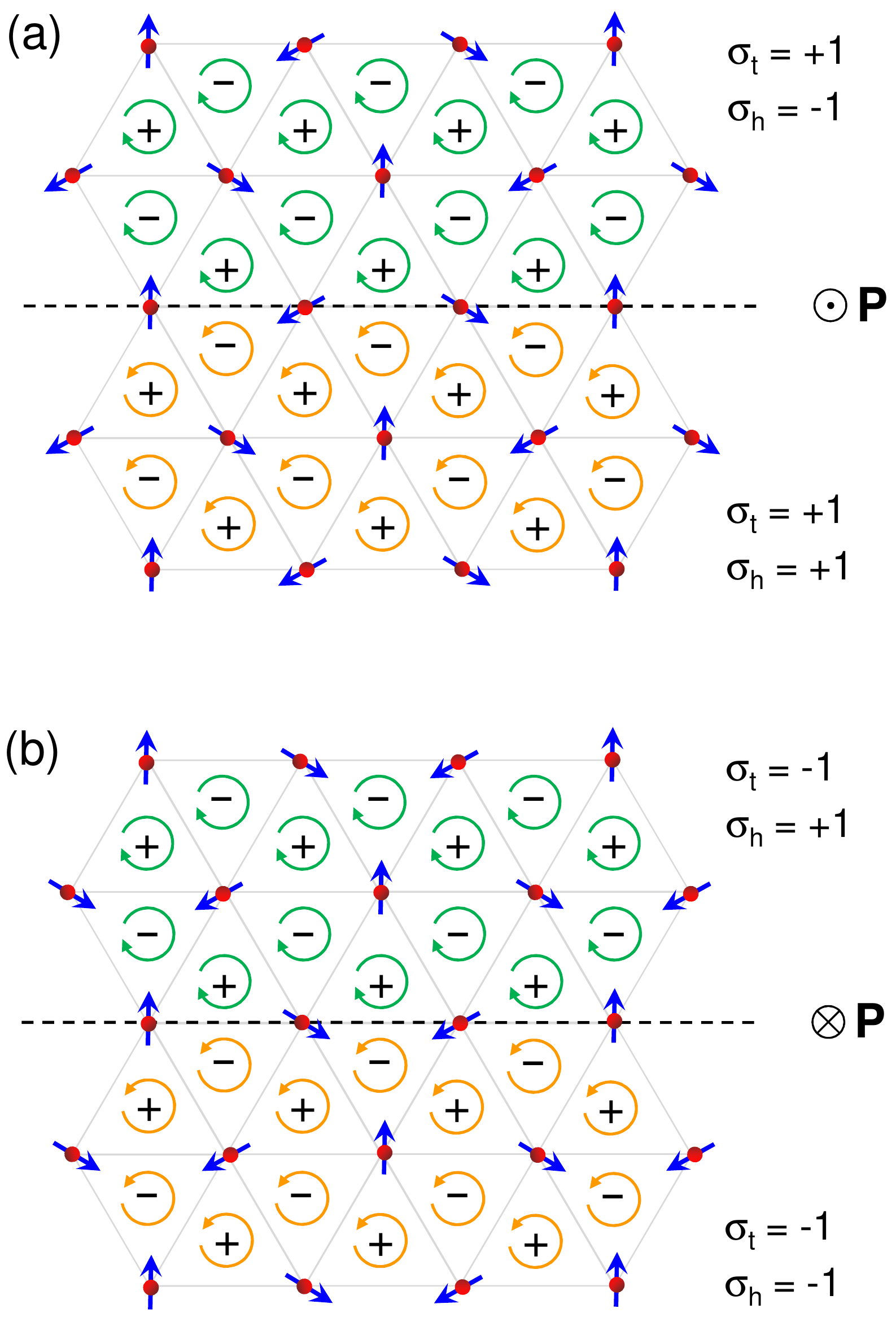}
\caption{(Color online) Magnetic structures present in (a) positive, and (b) negative field cooling, each of which has two contributions (from each axial domain). The direction of the ferroelectric polarisation ($\mathbf{P}$) is shown, and the axial distortion is indicted by the direction of the circular arrows. The direction of the MoO$_4$ tetrahedron associated with each triangle is shown by the $\pm$ signs. \label{Fig_domains}}
\end{center}
\end{figure}

RFMO single crystals were grown by a flux technique using high purity ($>99.9\%$) Rb$_2$CO$_3$, Fe$_2$O$_3$, and MoO$_3$ in a molar ratio of 2:1:6 according to the recipe described in \cite{waskowska2010temp_159}. They were heated together in air using a platinum crucible to 825$^\circ$C and kept at this temperature for 48 h. The homogenized melt was slowly cooled to 600$^\circ$C at a rate of 3$^\circ$C/h, followed by a faster rate of cooling to room temperature. Single crystal thin platelets (up to 1 cm in diameter) were separated from the flux, their quality and orientation were checked using an Agilent Technologies SuperNova diffractometer, and pyroelectric currents were measured and integrated to give the ferroelectic polarisation (along $c$) as a function of temperature (see Fig. \ref{Fig_Temp_dep}). Gold contacts were evaporated onto the $(0, 0, \pm1)$ surfaces and the OrientExpress neutron back-reflection Laue diffractometer \cite{Ouladdiaf2006orient_230} at the Institut Laue-Langevin (ILL), Grenoble, France was used to mount the sample with $\mathbf{a}^*+\mathbf{b}^*$ parallel to the $Y$-axis (see inset to Fig. \ref{Fig_Temp_dep}). This allowed us to access peaks of the form $(hhl)$. The contacts were connected to gold wires with silver epoxy to allow high voltage to be applied. An ILL `orange' cryostat provided cooling during the neutron scattering experiment (using a fixed neutron wavelength of 0.825 {\AA}), which was carried out using the CryoPad \cite{Tasset1999spherical_222} set-up on beamline D3 at the ILL.

Initially the sample was cooled below $T_\textrm{c}$, fixing the population of axial structural domains for the rest of the experiment. Fig. \ref{Fig_Temp_dep} shows the temperature dependence of the intensity of the $(1/3,1/3,0.44)$ magnetic peak (corresponding to the propagation vector $\mathbf{q}_1$), which fits the pyroelectric data well and confirms the simultaneous onset of magnetic ordering and ferroelectricity at $T_\textrm{N} \approx 4 $K. The same sample was cooled in an applied electric field of both $\pm$ 7.5 kV/cm and zero field, and the polarimetry components measured. The results for the $(-1/3,-1/3,0.56)$ peak are shown in Fig. \ref{Fig_Bars_Hyst}(a), together with the calculations (which contain no free parameters) from equations (\ref{pol1}) to (\ref{pol2}). It is clear that the components $P_{yx}$ and $P_{zx}$ couple to the electric field (slight differences in these components, which should be equal in magnitude, are due to experimental uncertainty). 

\begin{table}[t]
\begin{center}
\caption{The eight possible contributions of the magnetic structure to the scattered intensity. $\sigma_\textrm{t}$: the triangular chirality; $\sigma_\textrm{h}$: the helical chirality; $\mathbf{Q}$: the position of the peak in reciprocal space; $\mathbf{G}$: a reciprocal lattice vector; $\mathbf{q}_1 = (1/3,1/3,q_z)$; $\mathbf{q}_2 = (-1/3, -1/3, q_z)$. The magnetic structure factor $\mathbf{M}$ is given in terms of the orthonormal vectors $\hat{\mathbf{u}}$ and $\hat{\mathbf{v}}$.} \label{peaks}
\begin{tabular}{llllll}
\hline \hline
$\sigma_\textrm{t} = 1$	&	$\sigma_\textrm{h} = 1$,	&	$\mathbf{Q} = \mathbf{G} \pm \mathbf{q}_1$,	&	$\mathbf{M} \propto (\hat{\mathbf{u}} \mp i\hat{\mathbf{v}})$	\\
				&	$\sigma_\textrm{h} = -1$,	&	$\mathbf{Q} = \mathbf{G} \pm \mathbf{q}_2$,	&	$\mathbf{M} \propto (\hat{\mathbf{u}} \pm i\hat{\mathbf{v}})$	 \\
\hline
$\sigma_\textrm{t} = -1 $	&	$\sigma_\textrm{h} = 1$,	&	$\mathbf{Q} = \mathbf{G} \pm \mathbf{q}_2$,	&	$\mathbf{M} \propto (\hat{\mathbf{u}} \mp i\hat{\mathbf{v}})$	 \\
				&	$\sigma_\textrm{h} = -1$,	&	$\mathbf{Q} = \mathbf{G} \pm \mathbf{q}_1$,	&	$\mathbf{M} \propto (\hat{\mathbf{u}} \pm i\hat{\mathbf{v}})$	 \\
\hline \hline
\end{tabular} 
\end{center}
\end{table}

Table \ref{peaks} lists the magnetic peaks and associated magnetic structure factors originating from different chiral structures. Peaks with propagation vectors  $\mathbf{q}_1$ and $\mathbf{q}_2$ are present with similar intensities, yielding four satellites around each reciprocal lattice node;  this shows that the crystal contains a roughly equal population of both axial domains.  We examined the behaviour of several magnetic peaks, including at least one from each combination of chiralities, in each of the field coolings. In zero field cooling, the off-diagonal components $P_{yx}$ and $P_{zx}$ for all peaks are zero. This is only possible if the fractions of ($+,+$) and ($-,-$) domains (which have equal and opposite $\Im{\{\mathbf{M}\}}$) are equal, \textit{i.e.} $f_{+,+}= f_{-,-}=0.5$, and likewise $f_{+,-}= f_{-,+}=0.5$.  For \emph{positive} field cooling, the non-zero off-diagonal terms require that only two of the chirality combinations be present, so that $f_{+,+}=f_{+,-}=1$ and $f_{-,+}=f_{-,-}=0$ [Fig. \ref{Fig_domains}(a)].  For \emph{negative} field cooling, the other two domains are populated, so that  $f_{+,+}=f_{+,-}=0$ and $f_{-,+}=f_{-,-}=1$ [Fig. \ref{Fig_domains}(b)]. Fig. \ref{Fig_Bars_Hyst}(b) shows the hysteresis in $P_{zx}$ as a function of applied electric field for two magnetic peaks. The $(1/3, 1/3, 0.44)$ peak (resulting from one axial domain) switches between $(+,+)$ and $(-,-)$ at $\pm E$ whereas the $(-1/3,-1/3,0.44)$ peak (from the other axial domain) switches between $(+,-)$ and $(-,+)$. This confirms that we have direct control over the magnetic structures in both axial domains by using an applied $\mathbf{E}$-field.

Based on the geometry of the exchange interactions (Fig. \ref{Fig_structure}), we expect that $J_1 > J_2 > J_\textrm{b} > J_\textrm{a} $. We used band structure calculations within the LSDA+U approximation \cite{Anisimov-97} ($U=4.5$~eV and $J_H=1.0$~eV) and the linear muffin-tin orbitals method \cite{Andersen-84} to estimate values of the two largest interactions, $J_1$ and $J_2$. These exchange parameters were calculated as the second derivatives of the energy variation at small spin rotations \cite{Liechtenstein} and were found to be 1.0~K and 0.3~K, respectively. Using eq. (\ref{prop_vec}) under the assumption that $J_\textrm{b} \gg J_\textrm{a}$ for $\sigma_\textrm{t} \sigma_\textrm{h} = 1$, together with $q_z = 0.44$, we estimate that $J_\textrm{b}-J_\textrm{a} \approx J_2/3$, leading to a separation in energy between the same magnetic configuration in each of the two axial domains of 0.40 K per spin. Our measured ferroelectric polarisation implies that an electric field of 7.5 kV/cm separates $(+,+)$ and $(-,-)$ states [Fig. \ref{Fig_Bars_Hyst}(b)] by $4\times 10^{-4}$ K per spin, \textit{i.e.} 1/1000 of the energy associated with the axial distortion.  

Since the direction of the electrical polarisation $\mathbf{P} = (0,0,P_z)$, as deduced by the direction of $\mathbf{E}$, is coupled to the magnetic structure, from symmetry considerations the magnetoelectric free energy can be written as $S^2P_z(c_1 \sigma_\textrm{t} +c_2 A \sigma_\textrm{h})$ where the axial vector $\mathbf{A} = (0,0,A)$ and $c_{1,2}$ are constants \cite{footnote2}.  The two terms in  this expression are made proportional to each other by eqn. (\ref{energy}) (since the sign of $A$ determines the difference in $J_\textrm{a}$ and $J_\textrm{b}$) which imposes that $\sigma_\textrm{t} = - \frac{A}{|A|} \sigma_\textrm{h}$. Our results show that the overall triangular chirality of the entire field-cooled sample is uniform and switches with the electric field, as conjectured by Kenzelmann \textit{et al.} \cite{kenzelmann2007direct_160}, in spite of the axial domain structure. However, as explained above, reversing the electric field actually switches both $\sigma_\textrm{h}$ and $\sigma_\textrm{t}$ simultaneously within a single structural domain (Fig. \ref{Fig_domains}).  The two chiral coupling terms to  $\sigma_\textrm{t}$ and $A\sigma_\textrm{h}$ are both of relativistic origin, since, as we have already shown, domains with opposite polarisation have the same symmetric-exchange energy.  However, the two terms  rely on different microscopic mechanisms; since the second term is proportional to the axial rotation, a systematic study of isostructural compounds, through measurements or first-principle calculations, could reveal which of the two provides the dominant contribution to the development of the electrical polarisation.

We have demonstrated that the axial distortion of the crystal structure plays a crucial role in stabilising the helical magnetic structure of RbFe(MoO$_4$)$_2$, linking triangular chirality with helicity.  By means of neutron spherical polarimetery, we determined uniquely the populations of domains with each of the combinations of helical and triangular chiralities in the two axial domains. By applying an external electric field, the domain population  switches between the pairs of magnetic structures that are energetically preferred in each axial domain.  The coupling between magnetic structure and electrical polarisation is of relativistic origin, and involves both triangular chirality and helicity, the latter term being proportional to the axial distortion (ferroaxial coupling).


\begin{thebibliography}{10}
\bibitem{footnote2}
Note that terms of the form $P_z \sigma_\textrm{h}$ and $P_z A \sigma_\textrm{t}$ are forbidden because they are not invariant by the choice of the positive $c^*$ direction, which changes the sign of $\mathbf{A}$, $\mathbf{P}$, and $\sigma_\textrm{t}$, but leaves $\sigma_\textrm{h}$ unchanged.
\end{thebibliography}
\end{document}